\documentstyle[amssymb,aps,multicol,epsf]{revtex}

\begin{document}
\draft
\title{Critical phenomena in networks }
\author{A.V. Goltsev$^{1,2,*}$, S.N. Dorogovtsev$^{1,2,\dagger }$, and J.F.F. Mendes$%
^{1,3,\ddagger }$}
\address{
$^1$ Departamento de F\'{\i }sica and Centro de F\'{\i }sica do Porto,\\
Faculdade de Ci\^{e}ncias, Universidade do Porto \\
Rua do Campo Alegre 687, 4169-007 Porto, Portugal\\
$^2$ Ioffe Physico-Technical Institute, 194021 St. Petersburg, Russia\\
$^3$ Departamento de F\'{\i }sica, Universidade do Aveiro, Campus\\
Santiago,3810-193 Aveiro, Portugal}
\maketitle

\begin{abstract}
We develop a phenomenological theory of critical phenomena in networks with
an arbitrary distribution of connections $P(k)$. The theory shows that the
critical behavior depends in a crucial way on the form of $P(k)$ and differs
strongly from the standard mean-field behavior. The critical behavior
observed in 
various networks is analyzed and found to be in agreement with the theory.
\end{abstract}


\begin{multicols}{2}
\narrowtext


Many complex, interacting systems such as the brain, Internet, social
systems, etc. are recognized as networks. They demonstrate a spectrum of
unique effects \cite{ba99,s01,ab01a,dm01c,ajb00d}, which distinguish them 
from all other known structures of condensed matter. One could expect that
such systems have peculiar cooperative phenomena. Only the first studies of
particular cooperative models on particular networks were performed recently 
\cite
{mn00d,ahs01,dgm02,lvvz02,g00,bw00,khhjmc01,hkc02,b02,cebh00,cnsw00,cbh02,pv01}%
. They were focused on traditional models in statistical physics, such as
the Ising and $XY$ models \cite
{ahs01,dgm02,lvvz02,g00,bw00,khhjmc01,hkc02,b02}, 
percolation \cite{cebh00,cnsw00,cbh02}, epidemic spreading, et al. in
scale-free networks \cite{pv01}. It was revealed that their critical
behavior is more rich and extremely far from that expected from the standard
mean-field theory. A deviation from the usual mean-field critical behavior
appears when, depending on the model, the fourth, $\left\langle
k^{4}\right\rangle $, or third, $\left\langle k^{3}\right\rangle $, moments
of the degree distribution $P(k)$ diverge. Here, {\em degree }is the number
of connections of a vertex. When $\left\langle k^{2}\right\rangle $
diverges, all models undergo an unusual phase transition of 
infinite order. This case is of 
primary importance because many real networks such as the Internet and
biological nets are described by $P(k)$ with infinite $\left\langle
k^{2}\right\rangle $.

Why do critical phenomena in networks differ so 
much from those in usual substrates and what is their common origin? Why do
all investigated models demonstrate universal behavior when $\left\langle
k^{2}\right\rangle $ diverges? In order to answer the 
questions raised above and analyze results of simulations and experiments
from a general point of view, it is necessary to have a general theory which
is not restricted by specific properties of any model.

The advantage of the phenomenological approach is that the origin of
interactions and nature of interacting objects are irrelevant. 
These can be spins, percolating clusters, biological objects and etc. In
this Letter, we develop a phenomenological theory of cooperative phenomena
in networks. Our approach is based on concepts of the Landau theory of
continuous phase transitions. It is shown that the critical behavior in
networks has a universal character and is determined by two general
properties: (i) the structure of a network and (ii) the symmetry underlying
a model. We find that in networks described by a degree distribution with
divergent moments, the thermodynamic potential of an interacting system with
cooperative effects is a singular function of an order parameter. The theory
is in complete agreement with exact results for the Ising and Potts models
and percolation on `equilibrium' uncorrelated random networks. On the basis
of the theory we discuss results of theoretical studies and simulations of
the Ising and $XY$ models and epidemic spreading on evolving and small-world
networks.

Let us consider a system of interacting objects. Interactions or links
between these objects form a net. We assume that some kind of `order' can
emerge. This `ordered' phase may be characterized by some quantitative
characteristic $x$ while it will vanish in a `disordered' phase above a
critical point. 
In order to study the critical behavior we assume that the thermodynamic
potential $\Phi $ of the system is not only a function of the order
parameter $x$ but also depends on the degree distribution: 
\begin{equation}
\Phi (x,H)=-Hx+\sum_{k}^{\infty }P(k)\phi (x,kx)
\,.  
\label{assump0}
\end{equation}
Here $H$ is a field conjugated with $x$. It should be emphasized that Eq. (%
\ref{assump0}) is not obvious {\it a priori}. The function $\phi (x,kx)$ may
be considered as a contribution of vertices with $k$ connections. There are
arguments in favor of this assumption. Let us consider the interaction of an
arbitrary vertex $0$ with $k$ neighboring vertices. In the framework of a
mean-field approach, $k$ neighboring sites having a spontaneous `moment' $x$
produce an effective field $kx$ acting on the site $0$. This indicates that
the expansion is over not only $x$ but also $kx$. 

We assume that 
$\phi(x,y)$ is a smooth function of $x$ and $y$ and can be represented as a
series in powers of both $x$ and $y$: 
\begin{equation}
\phi(x,y) = \sum_{m=0}^{\infty}\sum_{l=0}^{\infty}\phi_{ml}x^m y^l 
\, ,
\label{assump1}
\end{equation}
where $\phi_{ml}$ are functions of `temperature' $T$ and $H$. We use the
term `temperature' for convenience. If a change of another parameter leads
to the emergence of `order', then this control parameter can be used as $T$.

There are general restrictions on the coefficients $\phi_{ml}$. For an
arbitrary $P(k)$, at zero field $H=0$, the expansion of $\Phi(x,H=0)$ over $%
x $ must contain only the second and higher powers of $x$. Therefore, $%
\phi_{00}=\phi_{01}=\phi_{10}=0$.

$\Phi $ must be finite if $\left\langle k\right\rangle $ is finite. This
condition is satisfied if at $y\gg 1$ and an arbitrary $x$ the function $%
\phi (x,y)$ increases slower than $y$: 
\begin{equation}
\phi (x,y)\leqslant g(x)y 
\, ,  
\label{assump2}
\end{equation}
where $g(x)$ is a function of $x$.

Eqs. (\ref{assump0})--(\ref{assump2}) are the basis of our theory.

In the framework of the Landau theory, it is assumed that the thermodynamic
potential $\Phi$ near a phase transition point can be represented as a
series in powers of the order parameter $x$: 
\begin{equation}
\Phi(x,H) = -Hx + f_2 x^2 + f_3 x^3 + f_4 x^4 + \ldots 
\, .  
\label{e4}
\end{equation}
Higher terms in the expansion (\ref{e4}) are irrelevant for the critical
behavior. The coefficients $f_n$ are related to the coefficients $\phi_{ml}$
in the expansion (\ref{assump1}). 
Let us suppose first that the critical temperature $T_c$ is finite. (The
case of an `infinite' critical temperature will be considered separately.) 
Following Landau, we assume that near the critical temperature the
coefficient $f_2$ can be written as $a(T-T_c)$ where $a>0$ due to the
condition of stability of the disordered phase at $T>T_c$. $f_2=0$ at the
transition point. 
This gives the following equation for $T_c$: 
\begin{equation}
\phi_{20}(T_c) + \phi_{11}(T_c) \left\langle k \right\rangle +
\phi_{02}(T_c) \left\langle k^2 \right\rangle = 0 
\, .  
\label{crit}
\end{equation}
According to this equation, in the case $\phi_{20}=0$ the critical
temperature $T_c$ depends only on the ratio $\left\langle k^2\right\rangle
/\left\langle k\right\rangle$. Such a dependence was revealed in all studied
models on networks \cite{dgm02,lvvz02,cebh00,cbh02,pv01}. In the ordered
phase the order parameter $x $ is determined by the condition that $\Phi
(x,H)$ is minimum: 
\begin{equation}
d\Phi (x,H)/dx=0 
\, .  
\label{eos}
\end{equation}
Solving Eq. (\ref{eos}), we find $x(T,H)$ and 
all other physical parameters: the response function $\chi _x=dx/dH$, the
specific heat $C=-Td^2\Phi /dT^2$, et. al. 

The condition of the stability of the ordered phase near the 
continuous phase transition demands that either $\left\langle
f_{3}\right\rangle >0$ or if $\left\langle f_{3}\right\rangle =0,$ then $%
\left\langle f_{4}\right\rangle >0$. In the general case, $f_{2n+1}\neq 0$.
If for 
symmetry reasons $\Phi (x,H)=\Phi (-x,-H)$, then $f_{2n+1}=0$. This
condition takes place if $\phi_{ml}=0$ at $m+l = 2n+1$.

Let us consider analytical properties of $\Phi$ in the general case. The $n$%
-th derivative of $\Phi (x,H)$ at $x=0$ is equal to 
\begin{equation}
\Phi^{(n)}(0)\equiv \left. d^{n}\Phi (x,H)/dx^{n}\right| _{x=0} = n!
\sum_{l=0}^{n} \phi_{n-l,l}\left\langle k^{l}\right\rangle 
\, .
\label{derriv}
\end{equation}
If all the moments $\left\langle k^{l}\right\rangle =\sum_{k}P(k)k^{l}$
converge, then all the derivatives $\Phi^{(n)}(0)$ are finite. Therefore, $%
\Phi (x,H)$ is a smooth function of $x$, at least at small $x$. However, in
most interesting real networks, $P(k)$ has divergent moments. This leads to
a very important consequence. If moments $\left\langle k^{l}\right\rangle $
with $l<p$ converge and moments $\left\langle k^{l}\right\rangle $ with $%
l\geqslant $ $p$ diverge, then $\Phi^{(l)}(0)$ with $l<p$ are finite while $%
\Phi ^{(l)}(0)\rightarrow \infty $ for $l\geqslant p$. In this case $\Phi
(x,H)$ is a singular function of $x$ and can be written as follows: 
\begin{equation}
\Phi (x,H) = -Hx + \sum_{n=2}^{p-1} f_{n}(k) x^{n} + x^{p}s(x) 
\, ,
\label{sFL}
\end{equation}
where $s(x)$ is a singular function: $s(0),$ $s^{(n)}(0)\rightarrow \infty $%
, but $xs(x)\rightarrow 0$ at $x\rightarrow 0$. This result is of primary
importance. It is the singular function that can lead to a deviation from
the standard mean-field behavior. A method for determining $s(x)$ is
proposed below. For convenience we use a power-law degree distribution $%
P(k)\propto k^{-\gamma }$. Then, $\langle k^{4}\rangle $ diverges for $%
\gamma \leq 5$, $\langle k^{3}\rangle $ diverges for $\gamma \leq 4$, and $%
\langle k^{2}\rangle $ is divergent for $\gamma \leq 3$.

We begin with the case $f_{2n+1}=0$. Let us consider the critical behavior
for $f_4 > 0$ and different $\gamma$.

(a) $\gamma >5$. Here, $\left\langle k^4\right\rangle $ converges. The
coefficient $f_4$ is finite and $\Phi $ has a usual form (\ref{e4}). At $H=0$
Eq. (\ref{eos}) leads to the standard critical behavior: 
\begin{equation}
x\propto \tau ^{1/2}\text{, }\Delta C\neq 0\text{, }\chi _x(H=0)\propto \tau
^{-1}.\text{ }  \label{results-mf1}
\end{equation}
Here, $\tau =1-T/T_c$, $\Delta C\sim 1/\left\langle k^4\right\rangle $ is
the jump of the specific heat at $T_c$. It tends to zero when $\gamma
\rightarrow 5$.

(b) $3<\gamma \leqslant 5$. Here, $\langle k^{4}\rangle $ diverges.
According to Eq. (\ref{sFL}), a singular term $x^{4}s(x)$ is expected to
appear in $\Phi $. The function $\phi (x,y)$ in Eq. (\ref{assump1}) may be
divided into two parts: 
\begin{equation}
\phi (x,y)=\sum_{m=0}^{\infty }\,\,\sum_{1\leqslant l<\gamma -1}\!\phi
_{ml}x^{m}y^{l}+G(x,y)
\,.  
\label{e10}
\end{equation}
The first term determines a non-singular contribution to $\Phi $ in Eq. (\ref
{assump0}). The second term, $G(x,y)$, gives the singular contribution. 
For $4<\gamma \leqslant 5$ the function $G(x,y)$ has the following properties: $%
G(0,y)\propto y^{4}$ at small $y$, $d^{n}G(0,y)/dy^{n}\neq 0$ for $n\geq 4$
at $y=0$. Owing to the condition (\ref{assump2}), the function $G(x,y)$
increases slower than $y^{3}$ at $y\gg 1$.

It is convenient to use a continuous approximation for a degree distribution 
$P(k)=Ak^{-\gamma }$, where $A$ is a normalization factor. Then Eq. (\ref
{assump0}) may be rewritten in the following form: 
\begin{equation}
\Phi (x,0)=f_{2}x^{2}+\varphi (x)x^{4}+A\int_{m}^{\infty }\frac{dk}{%
k^{\gamma }}G(x,xk)
\,,  
\label{FL-3}
\end{equation}
where $m$ is the smallest degree in $P(k)$, $\varphi (x)$ is a smooth
function which is determined by the convergent moments $\langle k^{l}\rangle 
$ with $l\leqslant 3$, $\varphi (0)\neq 0$. In order to find $\Phi $ at
small $x$, one can put $G(x,y)\approx G(0,y)$ into Eq. (\ref{FL-3}).
Considering the integral over a variable $y=xk$, one can show that the
region $mx\leq y\lesssim b$ gives a leading contribution. 
$b$ is a model parameter. We obtain 
\begin{eqnarray}
\Phi (x,H) &=&-Hx+a(T-T_{c})x^{2}+Ax^{4}\ln (b/x)
\,,  
\label{Free-5} 
\\[5pt]
\Phi (x,H) &=&-Hx+a(T-T_{c})x^{2}+Bx^{\gamma -1}
\,,  
\label{Free-3-5}
\end{eqnarray} 
for $\gamma =5$ and $3<\gamma <5$, respectively. $A$ and $B$ are constants
at $T=T_{c}$. The critical behavior for the case $f_{3}=0$, $f_{4}>0$ at $%
\gamma >3$ is summarized in Table \ref{t.1}. It was observed in the Ising
model on `equilibrium' uncorrelated random networks \cite{dgm02,lvvz02}.

Note that the divergence of $\langle k^{4}\rangle$ does not change the
critical behavior of $\chi _{x}$. 

(c) $2<\gamma \leqslant 3$. Now $\left\langle k^{2}\right\rangle $
diverges. 
Equation (\ref{crit}) shows that in this situation there is no phase
transition at any finite temperature. 
According to Eq. (\ref{sFL}), a singular term $x^{2}s(x)$ is expected to
appear in $\Phi $. Calculations of $\Phi $ may be carried out in a similar
way as above, using the function $G(x,y)=\phi (x,y)-\sum_{m=1}\phi
_{m1}x^{m}y$, Eq. (\ref{e10}). We find 
\begin{eqnarray}
\Phi (x,H) &=&-Hx+Cx^{2}-Dx^{2}\ln (b/x)\,,  \label{Free-3} \\[5pt]
\Phi (x,H) &=&-Hx+C^{\prime }x^{2}-D^{\prime }x^{\gamma -1}  \label{Free-2-3}
\end{eqnarray}
for $\gamma =3$ and $2<\gamma <3$, respectively. The coefficients $C,$ $D$ , 
$C^{\prime }$ and $D^{\prime }$ are functions of $T$ and $H$. If $C,$ $D$ , $%
C^{\prime }$ and $D^{\prime }$ are positive, then at arbitrary finite
temperature $T\gg 1$ in zero field $H=0$, Eq. (\ref{eos}) has stable
nontrivial solutions: 
\begin{eqnarray}
x &=&b\exp (-(2C+D)/(2D)),\text{ }\chi _{x}=1/(2D)
\,,  
\label{results-3} 
\\[5pt]
x &=&\left( \frac{2C^{\prime }}{(\gamma -1)D^{\prime }}\right)
^{\!-1/(3-\gamma )},\text{ }\chi _{x}=\frac{1}{2(3-\gamma )C^{\prime }}
\label{results-2-3}
\end{eqnarray}
for $\gamma =3$ and $2<\gamma <3$, respectively. 
As in the Landau theory, the results (\ref{results-3}) and (\ref{results-2-3}%
) are obtained in terms of the coefficients of the thermodynamic potential.
Note that in the situation where a phase transition is absent at any finite
temperature, 
the phenomenological theory can not determine the temperature behavior of
the coefficients $C,$ $D$, $C^{\prime }$ and $D^{\prime }$. 
In this situation, the temperature dependences of the coefficients can be
found only by a microscopic theory (see below).

In the general case the symmetry of the model 
admits non-zero $f_{3}$. In the case $f_{3}>0$ the analysis of the
analytical properties of $\Phi $ can be performed as above. For $\gamma >4$, 
$\Phi $ is a smooth function of $x$ and leads to the standard critical
behavior. At $\gamma \leqslant 4$, $\Phi $ contains a singular term: (a) $%
x^{3}\ln x$ at $\gamma =4;$ (b) $x^{\gamma -1}$ in the ranges $4>\gamma >3$
and $3>\gamma >2$; (c) $x^{2}\ln x$ at $\gamma =3.$ The corresponding
critical behavior for $\gamma >3$ is represented in Table \ref{t.1}. This
behavior was observed for percolation on 
uncorrelated random networks \cite{cbh02}. In the range $2<\gamma \leqslant
3 $ when $\left\langle k^{2}\right\rangle $ diverges, $\Phi $ has the
universal form, Eqs. (\ref{Free-3}) and (\ref{Free-2-3}).

In the case 
$f_{3}<0$, or $f_{4}<0$\ if $f_{3}=0$, when $\left\langle k^{2}\right\rangle 
$ converges, a first-order phase transition occurs. In agreement with this
prediction we found such a transition in the $q-$state Potts model with $%
q\geqslant 3$ on `equilibrium' uncorrelated random networks by use of the
approach of Ref. \cite{dgm02}. In the limit $\gamma \rightarrow 3$ the jump
of the order parameter at the transition tends to zero, and the transition
transforms into the infinite order phase transition. A detailed study of the
transition will be given elsewhere.

In order to complete Table \ref{t.1}, let us discuss the temperature
behavior in the case $2<\gamma \leqslant 3$ within the microscopic theory of
the Ising model and percolation on `equilibrium' uncorrelated random
networks \cite{ahs01,dgm02,lvvz02,cbh02}. For this purpose we use the more
general ferromagnetic $q-$state Potts model which at $q=1$ and 2 is
equivalent to percolation and the Ising model, respectively (see, for
example Ref. \cite{wu82}). Using the approach of Ref. \cite{dgm02} we obtain
that in the Potts model a continuous phase transition occurs at the exact
critical temperature 
\begin{equation}
T_{c}=2/\ln \frac{\langle k^{2}\rangle +(q-2)\left\langle k\right\rangle }{%
\langle k^{2}\rangle -2\langle k\rangle } 
\,.  
\label{Tc}
\end{equation}
Hereafter, we set the energy of ferromagnetic interaction between nearest
neighbors $J=1$. The parameter $p_{c}=1-\exp (-2/T_{c})$ determines the
percolation threshold \cite{cnsw00,cbh02} and establishes the relation
between the Potts model and percolation.

Let us consider the case $2<\gamma \leqslant 3.$ Here, $\left\langle
k^{2}\right\rangle $ diverges. $T_{c}$ is infinite for the infinite
networks. In any finite network, $\left\langle k^{2}\right\rangle <\infty $,
and $T_{c}$ is finite, although it may be very high, $T_{c}\cong 2\langle
k^{2}\rangle /(\left\langle k\right\rangle q)$. In the temperature region $%
T\gg 1$, where $x\ll 1$, the Potts model demonstrates universal behavior at
all $q\geqslant 1$: 
\begin{eqnarray}
x &\approx &(q/\langle k\rangle )e^{-qT/\left\langle k\right\rangle },\text{%
\quad }\chi _{x}\propto T^{-1}
\,,  
\label{gamma3} 
\\[5pt]
x &\sim &T^{-1/(3-\gamma )}\!,\text{ \quad }\chi _{x}\propto T^{-2}
\label{gamma-2-3}
\end{eqnarray}
for $\gamma =3$ and $2<\gamma <3$, respectively. Without the continuum
approximation, we have instead of $\langle k\rangle $ in the exponential, a
constant which is determined by the complete form of $P(k)$. In Ehrenfest's
classification, this transition is of infinite order, as all temperature
derivatives of $\Phi $ are finite. The results (\ref{results-3}) and (\ref
{results-2-3}) agree with Eqs. (\ref{gamma3}) and (\ref{gamma-2-3}) if we
put $C,C^{\prime }\propto T^{2}$, $D,D^{\prime }\propto T$. The exponential
behavior (\ref{gamma3}) has been revealed in epidemic spreading within
scale-free networks with $\gamma =3$ \cite{pv01} 
and in percolation on these networks \cite{cbh02}.

At small $x$ there is the following relationship between the response
function $\chi _{x}$ and the susceptibility $\chi =dM/dH$: $\chi \approx
2/qT+\left\langle k\right\rangle \chi _{x}/q$. At $\gamma >3$ the critical
behavior of $\chi $ is determined by $\chi _{x}$: $\chi \approx \left\langle
k\right\rangle \chi _{x}/q$. At $2<\gamma \leqslant 3$ we have $\chi \propto
1/qT,$ as the paramagnetic contribution $2/qT$ is 
of the order of $\chi _{x}$ or larger.

Let us discuss the results of theoretical studies and simulations of
critical phenomena in different networks on the basis of the
phenomenological theory.

The phenomenological theory as well as the Landau theory assumes that the
contribution of fluctuations to the thermodynamic potential is small. Above
we have shown that the theory gives the exact critical behavior of the Ising
model, percolation and Potts model on `equilibrium' uncorrelated random
networks \cite{dgm02,lvvz02,cbh02}. The reason for this is that these
networks have a local tree-like structure and vertices are statistically
equivalent\cite{bbk72,mr95,nsw00}. Due to these properties, vertices in the
networks can be regarded as forming a Bethe-lattice structure for which a
mean-field approach is exact \cite{bbook82}. It means that the fluctuation
contribution is negligibly small. Note that in a graph with a Cayley
tree-like structure, vertices are statistically inequivalent, and a
mean-field approach is valid only for vertices deep within such a graph \cite
{bbook82}.

The origin of a deviation from the standard mean-field critical behavior is
different for regular lattices and networks. In a regular lattice the
deviation is caused by strong critical fluctuations which depend crucially
on the space dimension. In networks the deviation is brought about by the
most connected vertices which induce strong correlations in their close
neighborhood. With decreasing 
exponent $\gamma$ 
the relative number of highly connected vertices increases and their role
turns out to be more important.

Recent simulations \cite{ahs01} of the Ising model on the
Barab\'{a}si-Albert scale-free network with the degree distribution $%
P(k)=Ak^{-3}$ revealed a temperature behavior described by Eq. (\ref{gamma3}%
) as well as in the Potts model on `equilibrium' uncorrelated random
networks. Unlike the latter networks, the Barab\'{a}si-Albert 
net is correlated. Nevertheless, results of the simulations agree with the
universal temperature behavior predicted by the phenomenological theory for
networks with the divergent moment $\left\langle k^{2}\right\rangle $.

The exact results for percolation on small-world networks were obtained in
Ref. \cite{mn00d}. The Ising and $XY$ models on small-world networks were
studied analytically and by use of Monte Carlo simulations\cite
{g00,bw00,khhjmc01,hkc02}. These networks, introduced by Watts and Strogatz 
\cite{ws98}, have a Poisson-like degree distribution and large clustering
coefficients. It was found that the critical behavior is characterized by
the standard mean-field critical exponents. This result is in agreement with
the prediction of the phenomenological theory that the standard mean-field
critical behavior should occur in networks described by a degree
distribution with convergent moments.

Thus, the analysis of critical behavior of the Ising, Potts and $XY$ models,
percolation and epidemic spreading on 
uncorrelated random, scale-free and small-word networks shows that the
critical behavior in networks depends crucially on the form of the degree
distribution and the symmetry underlying a model in a complete agreement
with the phenomenological theory. The studied networks differ in clustering
coefficients, degree correlations, etc. However, the situation looks like
these characteristics are not relevant for critical behavior. Their role in
the formation of the critical exponents is still unclear. Further
investigations in this topic are necessary. It would be 
interesting to find a network where critical behavior differs from a
mean-field one due to strong fluctuations. However, even in this case it is
expected that when $T$ 
tends to $T_{c}$, a temperature region of the mean-field critical behavior 
precedes a region of strong fluctuations.

Real systems like the Internet, WWW or biological networks have 
a network structure with a power-law degree distribution with exponent $%
\gamma$ below 3, see Refs. \cite{ab01a,dm01c}. The phenomenological theory
predicts for this case the power-law critical behavior (\ref{gamma-2-3}),
see also the last line in Table \ref{t.1}. This behavior agrees with an
empirical observation \cite{ajb00d} for the nd.edu domain of the WWW, where
the variations of the size of the giant component under random damage were
studied. This size and the number of damaged vertices play a role of the
order parameter $x$ and the control parameter, respectively.

In conclusion, the phenomenological theory shows that the deviation of the
critical behavior of interacting systems with a network structure from the
standard mean-field behavior emerges when a degree distribution has
divergent moments. The theory predicts different classes of critical
behavior in agreement with microscopic studies of the Ising and Potts
models, and percolation on `equilibrium' uncorrelated random networks. It
also agrees with results previously obtained for various 
models on small-world and evolving networks. The theory can easily be
generalized for models with a multicomponent order parameter and can give
useful insight into collective effects in different networks discussed in
connection with the Internet, biological networks, etc. Using 
this approach, one can also study the critical relaxation in models on
networks.

S.N.D and J.F.F.M. were partially supported by the project
POCTI/99/FIS/33141. A.G. acknowledges the support of the NATO program
OUTREACH. We also thank A.N. Samukhin for many useful discussions\vspace{4pt}%
. 

\noindent $^{*}$ E-mail address: goltsev@pop.ioffe.rssi.ru \newline
$^{\dagger }$ E-mail address: sdorogov@fc.up.pt \newline
$^{\ddagger }$ E-mail address: \vspace{-8pt} jfmendes@fc.up.pt

\end{multicols}


\widetext

\begin{table}[tbp]
\begin{tabular}{cccl}
& $x$ & $\delta C(T<T_c)$ & $\!\!\!\!$$\chi_x$ \\[3pt] \hline
&  &  &  \\[-8pt] 
\begin{tabular}{cc}
\begin{tabular}{c}
$\ f_3=0$,$\phantom{W}$ \\ 
$\ f_4 > 0$$\phantom{W}$%
\end{tabular}
& $\left\{\phantom{WW} 
\begin{tabular}{c}
$\gamma>5$$\phantom{\frac{\stackrel{|}{i}}{i}\!\!}$ \\[4pt] 
$\gamma = 5$ \\[4pt] 
$3< \gamma < 5$%
\end{tabular}
\right.$ \\[24pt] 
$\ f_3 > 0$$\phantom{W}$ & $\left\{\phantom{WW} 
\begin{tabular}{c}
$\gamma > 4$ \\[4pt] 
$\gamma = 4$ \\[4pt] 
$3< \gamma < 4$%
\end{tabular}
\right.$%
\end{tabular}
& 
\begin{tabular}{c}
$\tau^{1/2}$ $\phantom{\frac{\stackrel{|}{|^|}}{i}\!\!}$ \\[3pt] 
$\tau^{1/2}/(\ln \tau^{-1})^{1/2}$ \\[4pt] 
$\tau^{1/(\gamma-3)}$ \\[4pt] 
$\tau$ \\[4pt] 
$\tau/(\ln \tau^{-1})$ \\[5pt] 
$\tau^{1/(\gamma-3)}$%
\end{tabular}
& $\left. 
\begin{tabular}{c}
jump at $T_c$$\phantom{\frac{\stackrel{|}{i}}{|}\!\!}$ \\[5pt] 
$1/\ln\tau^{-1}$ \\[5pt] 
$\tau^{(5-\gamma)/(\gamma - 3)}$ \\[5pt] 
$\tau$ \\[4pt] 
$\tau /(\ln^2 \tau^{-1})$ \\[5pt] 
$\tau^{(5-\gamma)/(\gamma-3)}$%
\end{tabular}
\phantom{Ww.} \right\}\!\!\!\!\!\!\!\!\!\!\!\!\!\!\!$ & 
\begin{tabular}{c}
$\!\!\!\!\!\!$$\tau^{-1}$%
\end{tabular}
\\[48pt] 
\begin{tabular}{cc}
\begin{tabular}{c}
arbitrary$\phantom{w}$ \\ 
$f_3$ and $f_4$$\phantom{w}$%
\end{tabular}
& $\left\{\phantom{WW} 
\begin{tabular}{c}
$\gamma = 3$$\phantom{\frac{|}{i}}$ \\ 
$\!\!2<\gamma <3$$\phantom{\frac{\stackrel{|}{.}}{|}}$%
\end{tabular}
\right.$%
\end{tabular}
& 
\begin{tabular}{c}
$e^{-cT}$$\phantom{\frac{\stackrel{|}{.}}{|}}$ \\[1pt] 
$T^{-1/(3-\gamma)}$$\phantom{\frac{|}{|}}$%
\end{tabular}
& 
\begin{tabular}{c}
$T^2e^{-2cT}$$\phantom{\frac{\stackrel{|}{.}}{|}}$ \\[2pt] 
$T^{-(\gamma -1)/(3-\gamma)}$%
\end{tabular}
& 
\begin{tabular}{c}
$\phantom{\frac{\stackrel{i}{i}}{|}}\!\!\!\!\!\!\!\!\!\!\!$$T^{-1}$ \\[2pt] 
$\!\!\!\!\!\!\!$$T^{-2}$%
\end{tabular}
\\[11pt] 
&  &  & 
\end{tabular}
\caption{ Critical behavior of the order parameter $x$, the specific heat $%
\protect\delta C$, and the response function $\protect\chi _{x}$ of a model
on networks with a degree distribution $P(k)\sim k^{-\protect\gamma }$ for
various values of the coefficients $f_{3}$ and $f_{4}$, and exponent $%
\protect\gamma $. $\protect\tau \equiv 1-T/T_{c}$, $c$ is a constant which
is determined by the complete form of $P(k)$. In the case $f_{3}<0,$ or $%
f_{4}<0$ if $f_{3}=0$, at $\protect\gamma >3$, the system undergoes a
first-order phase transition. }
\label{t.1}
\end{table}



\begin{references}
\bibitem{ba99}  A.-L. Barab\'{a}si and R. Albert, Science {\bf 286}, 509 
(1999).

\bibitem{s01}  S.H. Strogatz, Nature {\bf 401}, 268 (2001).

\bibitem{ab01a}  R. Albert and A.-L. Barab\'{a}si, Rev. Mod. Phys. {\bf 74},
47 (2002).

\bibitem{dm01c}  S.N.~Dorogovtsev~and~J.F.F.~Mendes, 
Adv. Phys. {\bf 51}, 1079 (2002).

\bibitem{ajb00d}  R. Albert, H. Jeong, and A.-L. Barab\'{a}si, 
Nature {\bf 406}, 378 (2000).

\bibitem{mn00d}  C. Moore and M.E.J. Newman, Phys. Rev. E {\bf 61}, 5678

(2000); Phys. Rev. E {\bf 62}, 7059 (2000); M.E.J. Newman, I. Jensen, and
R.M. Ziff, Phys. Rev. E {\bf 65}, 021904 (2002).

\bibitem{ahs01}  A. Aleksiejuk, J.A. Holyst, and D. Stauffer, Physica A {\bf %
310}, 260 (2002).

\bibitem{dgm02}  S.N.~Dorogovtsev, A.V. Goltsev~and~J.F.F.~Mendes, Phys. 
Rev. E {\bf 66,} 016104 (2002).

\bibitem{lvvz02}  M. Leone, A. V\'{a}zquez, A. Vespignani, and R. Zecchina,
Eur. Phys. J. B {\bf 28}, 191 (2002)

\bibitem{g00}  M. Gitterman, J.Phys. A: Math. Gen. {\bf 33}, 8373 (2000).

\bibitem{bw00}  A. Barrat and M. Weigt, Eur. Phys. J. B {\bf 13}, 547 (2000).

\bibitem{khhjmc01}  B.J. Kim, H. Hong, P. Holme, G.S. Jeon, P.Minnhagen, and
M.Y.Choi, Phys. Rev. E {\bf 64}, 056135 (2001).

\bibitem{hkc02}  H. Hong, B.J. Kim, and M.Y.Choi, cond-mat/0204357.

\bibitem{b02}  G. Bianconi, cond-mat/0204455.

\bibitem{cebh00}  R. Cohen, K. Erez, D. ben-Avraham, and S. Havlin, Phys.
Rev. Lett. {\bf 85}, 4626 (2000).

\bibitem{cnsw00}  D.S. Callaway, M.E.J. Newman, S.H. Strogatz, and D.J.
Watts, Phys. Rev. Lett. {\bf 85}, 5468 (2000).

\bibitem{cbh02}  R. Cohen, D. ben-Avraham, and S. Havlin, Phys. Rev. E {\bf %
66}, 036113 (2002). 

\bibitem{pv01}  R. Pastor-Satorras, and A. Vespignani, Phys. Rev. Lett. {\bf %
86}, 3200 (2001); Phys. Rev. E {\bf 63,} 066117 (2001).

\bibitem{wu82}  F.Y. Wu, Rev. Mod. Phys., {\bf 54}, 235 (1982).

\bibitem{bbk72}  A. Bekessy, P. Bekessy, and J. Komlos, Stud. Sci. Math.

Hungar. {\bf 7}, 343 (1972); E.A. Bender and E.R. Canfield, J. Combinatorial
Theory A {\bf 24}, 296 (1978); B. Bollob\'{a}s, Eur. J. Comb. {\bf 1}, 311
(1980); N.C. Wormald, J. Combinatorial Theory B {\bf 31}, 156,168 (1981).

\bibitem{mr95}  M. Molloy and B. Reed, Random Structures and Algorithms {\bf %
6}, 161 (1995). 

\bibitem{nsw00}  M.E.J. Newman, S.H. Strogatz, and D.J. Watts, Phys. Rev. E 
{\bf 64}, 026118 (2001).%

\bibitem{bbook82}  R.J. Baxter, {\em Exactly Solved Models in Statistical
Mechanics} (Academic Press, London, 1982).

\bibitem{ws98}  D.J. Watts and S.H. Strogatz, Nature {\bf 393}, 440 (1998).
\end{references}
\end{document}